# The static structure factor of amorphous silicon and vitreous silica

Adam M. R. de Graff and M. F. Thorpe[*]

*Physics Department, Arizona State University, Tempe, AZ 85287, USA.*
*E-mail: mft@asu.edu*

**Synopsis** The structure factors for amorphous silicon and vitreous silica in the static limit are determined from large computer models and compared with available experiments.

**Abstract** Liquids are in thermal equilibrium and have a non-zero static structure factor $S(Q \to 0) = [\langle N^2 \rangle - \langle N \rangle^2]/\langle N \rangle = \rho_0 k_B T \chi_T$ where $\rho_0$ is the number density, $T$ is the temperature, $Q$ is the scattering vector and $\chi_T$ is the isothermal compressibility. The first part of this result involving the number $N$ (or density) fluctuations is a purely geometrical result and does *not* involve any assumptions about thermal equilibrium or ergodicity and so is obeyed by all materials. From a large computer model of amorphous silicon, local number fluctuations extrapolate to give $S(0) = 0.035 \pm 0.001$. The same computation on a large model of vitreous silica using *only* the silicon atoms and rescaling the distances gives $S(0) = 0.039 \pm 0.001$, which suggests that this numerical result is robust and similar for all amorphous tetrahedral networks. For vitreous silica, we find that $S(0) = 0.116 \pm 0.003$, close to the experimental value of $S(0) = 0.0900 \pm 0.0048$ obtained recently by small angle neutron scattering. More detailed experimental and modelling studies are needed to determine the relationship between the fictive temperature and structure.

## 1. Introduction

Correlated density fluctuations over large length scales can be determined from the small $Q$ limit of the static structure factor $S(Q)$, and thus can be obtained directly from diffraction experiments (Egami & Billinge, 2003). The structure factor can be defined in terms of the real-space pair density $\rho(r)$ via the sine Fourier transform

$$Q[S(Q) - 1] = \int_0^\infty 4\pi r[\rho(r) - \rho_0] \sin Qr \, dr$$
$$= \int_0^\infty G(r) \sin Qr \, dr \quad (1)$$

where $\rho_0$ is the average density and $G(r) = 4\pi r[\rho(r) - \rho_0]$ is the pair distribution function. This is also a convenient way to obtain $S(Q)$ from computer generated structural models, as $\rho(r)$ and hence $G(r)$ is rather straightforward to compute.

Of interest here is the structure factor (Egami & Billinge, 2003) in the small $Q$ (corresponding to large distances) limit, $S(Q \to 0)$, which has rarely been discussed in the context of amorphous modelling but which is of considerable interest. We will refer to this limit as the static structure factor, which can be measured by small angle elastic scattering (i.e. diffraction) experiments using either x-rays or neutrons, and it is of considerable interest theoretically as it contains information about how far the system is from thermal equilibrium, which will be discussed later. In order to obtain any kind of reliable estimate of $S(0)$ from computer generated models, it is necessary for the model to be large, and in this paper we focus on the excellent models of amorphous silicon and vitreous silica developed by Mousseau, Barkema and Vink ( Barkema & Mousseau, 2000, Vink *et al.*, 2001, Vink & Barkema, 2003), which we will show are large enough so that a reliable estimate for $S(0)$ can be extracted.

From general considerations (Hansen & McDonald, 1986), there is a sum rule relating the limit $S(Q \to 0)$ to the variance in the number of atoms $N$ within a volume $V$, namely

$$S(0) = [\langle N^2 \rangle - \langle N \rangle^2]/\langle N \rangle \qquad (2)$$

in the thermodynamic limit as $V \to \infty$. In this paper, we demonstrate that the static structure factor in the small $Q$ limit is small but non-zero for realistic and large enough models of amorphous silicon and vitreous silica that numerical values can be obtained with some confidence. For crystals, with no variance in the density due to their periodicity, equation (2) gives $S(0) = 0$. Note that there are no assumptions about thermal equilibrium in the derivation of equation (2) which is of purely geometrical origin (Torquato & Stillinger, 2003).

If further assumptions about thermal equilibrium and ergodicity are made, there is the additional result, well known in liquid theory (Hansen & McDonald, 1986), that relates number fluctuations to the isothermal compressibility $\chi_T$, namely

$$[\langle N^2 \rangle - \langle N \rangle^2]/\langle N \rangle = \rho_0 k_B T \chi_T \,. \qquad (3)$$

This relation assumes that all the states of a system at temperature $T$ governed by a potential are sampled according to Boltzmann statistics. Hence for liquids (and other thermodynamic, ergodic systems in thermal equilibrium), we have

$$S(0) = \rho_0 k_B T \chi_T. \qquad (4)$$

Equation (4) is also true for multi-component systems if $\rho_0$ is interpreted as the atomic number density and $S(0) = S_{NN}(0)$ is a Bhatia-Thornton structure factor (Bhatia & Thornton, 1970, Salmon, 2006, 2007), where $N$ refers to the total number of atoms.

## 1.1. Amorphous materials

*Amorphous silicon* is perhaps the furthest from equilibrium of all amorphous materials. This is because it is highly strained, with most of the strain being taken up by deviations of the bond angles from their ideal tetrahedral value of 109.5°. Each silicon atom has 3 degrees of freedom. The important terms in the potential are the bond stretching and angle bending forces around each atom. There are 4 covalent bonds at each silicon atom, each of which is shared, giving a net of 2 bond stretching constraints per atom. Of the 6 angles at each silicon atom, 5 are independent, giving a total of 7 constraints per atom. As there are considerably more constraints than degrees of freedom, the network is highly over-constrained (Thorpe, 1983). In thermal equilibrium, silicon cycles between crystalline solid and liquid forms. There is no glass transition. However, amorphous silicon can be prepared by various techniques involving very fast cooling and provides an extreme example of a non-equilibrium state.

*Vitreous silica* is a bulk glass, which contains very little strain, as can be seen as follows. The important constraints are the bond stretching and angle bending forces associated with the silicon atoms as in amorphous silicon. The angular forces at the oxygen ions are weak (Sartbaeva *et al.*, 2006). The total number of constraints per $SiO_2$ unit is 4 Si-O bond stretching constraints plus 5 angular forces at the Si giving a total of 9 constraints. However, the number of degrees of freedom per $SiO_2$ is also 9 (3 per atom). The system is therefore isostatic and not over-constrained (Thorpe, 1983). Thus, the strong Si-O bond stretching and O-Si-O angle bending forces are well accommodated (although the weaker angular distribution at the oxygen atom less so), so that vitreous silica is closer to thermal equilibrium than amorphous silicon, although not close enough that equation (4) can be used. However, equation (4) is much more likely to be obeyed, *if* the fixative temperature $T_f$ at which the glass was formed is used instead of $T$ (including for the compressibility). A much slower decrease in $S(0)$ is observed as the temperature is decreased below $T_f$ due to the freezing out of thermal vibrations about a fixed topology, as shown in the extensive and informative experiments of Levelut et. al (Levelut *et al.*, 2002, Levelut *et al.*, 2005, Levelut *et al.*, 2007).

**1.2. Computer models**

There are a number of high-quality periodic computer generated models for amorphous silicon. The first set of coordinates is from a small model with 4096 atoms (henceforth called the 4096 atom model) (Djordjevic *et al.*, 1995), built within a cubic super-cell with sides of length $L$ = 43.42Å. The average bond length is $a$ = 2.35Å, equal to the known value for crystalline silicon, and the model has the same density as crystalline silicon, which is about right for structurally good samples of amorphous silicon containing few voids, defects, etc. The network was constructed using the WWW technique (Wooten *et al.*, 1985, Djordjevic *et al.*, 1995), based on locally restructuring the topology of crystalline silicon, while keeping the number of atoms and covalent bonds fixed, until the ring statistics settle down and there are no Bragg peaks apparent in the diffraction pattern.

The second model contains 100,000 atoms (referred to as the 100K model) within a cubic super-cell of sides $L$ = 124.05Å, with an average bond length of $a$ = 2.31Å, and was built using a modified WWW technique (Vink *et al.*, 2001) based on previous work by Barkema and Mousseau (Barkema & Mousseau, 2000). We note that the models of Mousseau and Barkema have the narrowest angular variance ( ~9°) at the silicon atoms ever achieved in a non-crystalline tetrahedral network, and they also avoid the issue of possible crystal memory effects in WWW type models, as they use a non-crystalline atomic arrangement initially. In this paper we use the 100K model used as a scaffold in modelling vitreous silica (Vink & Barkema, 2003). The 100K model, like other models built by Barkema and Mousseau (Barkema & Mousseau, 2000), has a density ~5% above that of crystalline silicon, which is too large for amorphous silicon. The reason why this model has a higher density, while being excellent in other aspects is not entirely clear, but it may be necessary to let the angular variance increase back up to ~ 11° in order to get the experimental density of amorphous silicon.  The correlation between this angular spread and the density needs further study. The increase in density is probably caused by the Keating potential used not being quite up to this level of sophistication. This difference should not affect the limit $S(Q \to 0)$ to first order, as an isotropic compression or expansion of the whole structure leaves the relative number fluctuations invariant in the thermodynamic limit.

A very large model of vitreous silica (300K model) has been produced by the same group (Vink & Barkema, 2003) by first decorating the 100K amorphous silicon model with an oxygen ion between each silicon ion  and relaxing appropriately. The covalent bond network was then modified using the WWW technique. With only a few exceptions, all silicon atoms maintain only oxygen atoms as covalently bonded neighbours and vice versa. An important difference between the 100K amorphous silicon and the 300K vitreous silica models is that by effectively changing the fundamental unit from a silicon atom to a corner sharing $SiO_2$ tetrahedron, the system is no longer overconstrained but instead isostatic (Thorpe, 1983), a point that was discussed in Section 1.1. One might expect the greater number of degrees of freedom and the lower internal stress of the vitreous silica model to affect the static structure factor, as vitreous silica is closer than amorphous silicon to thermal equilibrium. We will return to this point later.

## 2. Calculation of the structure factor in the limit $Q \to 0$

### 2.1. Directly from the set of pair separations

The static structure factor $S(Q)$ can be calculated in a number of ways, some of which are more useful (i.e. smoother) than others when extrapolating to $Q \to 0$. We focus first on amorphous silicon, a material with a single atomic species. The structure factor can be computed directly from the set of atom coordinates by taking the spherical average of

$$S(\mathbf{Q}) = 1 + \frac{1}{N\langle f \rangle^2} \sum_{i \neq j} f_j^* f_i \exp(i\mathbf{Q} \cdot \mathbf{r}_{ij}) \tag{5}$$

where $f_i$ is the scattering factor of atom $i$. A spherical average yields

$$S(Q) = 1 + \frac{1}{N\langle f \rangle^2} \sum_{i \neq j} f_j^* f_i \frac{\sin Q r_{ij}}{Q r_{ij}} \tag{6}$$

where the sum $i \neq j$ goes over all pairs of atoms (excluding the self terms) in the periodic cubic super-cell of size $L$, and is evaluated at $Q_{lmn} = \frac{2\pi}{L}\sqrt{l^2 + m^2 + n^2}$ where $l$, $m$, and $n$ are integers. For a finite model with periodic boundary conditions, this means that it does not matter if the distances $r_{ij}$ are measured *within* the unit super-cell or *across* unit super-cells, as long as all $N(N-1)$ terms are computed in equation (6).

This computational approach using equation (6) suffers from two problems. The first is that there are of order $N^2$ terms in the sum, which becomes computationally demanding for large models. Secondly, there are finite size effects at small $Q$, even with periodic boundary conditions, creating a peak in $S(Q)$ at the origin of finite width $\sim 1/L$ and amplitude $N$. The peak at small $Q$, studied by small angle x-ray or neutron scattering, is given by the convolution of the delta function that would exist at the origin if the model were infinite, with a function related to the shape of the box in which the model exists (Lei *et al.*, 2009). This problem at small $Q$ could in principle be alleviated by subtracting the peak at the origin due to the finite size of the model (or sample), but the form of the peak is only known algebraically for a limited set of shapes (Lei *et al.*, 2009) which do not include the cube for which a double angular integration is needed. The numerical subtraction of two large numbers $O(N)$ would lead to errors $O(1)$, which is the order of the answer required. A better approach to finding the form of $S(Q)$ in a form suitable for extrapolation to small $Q$ is described below. We note that it is $S(Q)$ in the limit as $Q \rightarrow 0$ that is of interest, and not $S(0)$ itself, as $Q = 0$ is a singular point.

**2.2. Fourier transform approach**

As a way to circumvent issues associated with the finite size of the sample that affect small $Q$, the structure factor $S(Q)$ can be obtained from $G(r)$ via the sine Fourier transform given in equation (1). It appears from the form of equation (1) as though the limit $S(Q \rightarrow 0)$ depends upon the sine transform of $G(r)$ alone, and thus the behavior of $G(r)$ at large $r$ does not contribute much to the limit $S(Q \rightarrow 0)$ [see Fig. 2 for an example of $G(r)$]. This can be shown to be false by expanding equation (1) in powers of $Q$ and keeping only the lowest order terms that would dominate in the small $Q$ limit. To the lowest order in $Q$

$$S(0) = 1 + \int_0^\infty 4\pi r^2 [\rho(r) - \rho_0] dr = 1 + \int_0^\infty r G(r)\, dr \qquad (7)$$

which depends on the integral of $rG(r)$, not $G(r)$. This factor of $r$ increases the sensitivity of $S(Q \rightarrow 0)$ to the details of the decay in $G(r)$ at large distances. Oscillations in $G(r)$ associated with a single reference atom are known to persist out to large distances (Levashov *et al.*, 2005) and are a serious concern when computing $S(Q \rightarrow 0)$ from a model. In practice, the use of equation (7) to find the limit $S(Q \rightarrow 0)$ also suffers from poor convergence at small $Q$, as $rG(r)$ amplifies the ripples that exist because of the finite nature of the model, although it is superior to using equation (6).

**2.3. Sampling volume method**

Quite generally, even in the absence of thermal equilibrium, the small $Q$ limit $S(Q \rightarrow 0)$ is related in the thermodynamic limit to number (or density) fluctuations within sub-regions of volume $V$ according to equation (2). As we only have models of finite size, even with periodic boundary conditions it is not possible to determine the limit directly and it is necessary to extrapolate to the $N \rightarrow \infty$ limit as best we can. The approach of extrapolating $S(Q)$ as $Q \rightarrow 0$ suffers from finite size effects that cause oscillations about the ideal $S(Q)$ which would be obtained for an infinite model. It is difficult to disentangle the finite size effects from the underlying ideal $S(Q)$, making accurate extrapolation always challenging.

A more accurate determination of $S(Q \to 0)$ can be achieved through equation (2). The equality states that the relative variance in the number of atoms within an ensemble of randomly placed, bounded, convex volumes (Torquato & Stillinger, 2003) is equal to $S(0)$ in the limit that the sampling volume goes to infinity. For a finite sampling volume of fixed shape, the variance in the number of atoms within the enclosed volume, which samples all possible positions and orientations equally, can be divided into terms that scale as the volume, those that scale as surface area, and those with lower order dependencies on the length scale of the enclosed volume (Torquato & Stillinger, 2003). If $R$ describes such a sampling length scale, then the relative variance, which divides the variance by the average number of atoms within the sampling volume, can be expressed as the sum of a volume term of order $R^0$, a surface term of order $R^{-1}$, and lower order terms.

Atomic structures for which the number variance does not depend on volume are called *hyperuniform*, examples of which are materials with a periodic lattice, as their unit cells have well defined volume and density. The number variance for such systems is related to the Gauss circle problem (Bleher *et al.*, 1993, Levashov *et al.*, 2005). The static structure factor for crystals is zero, as the structure factor $S(Q)$ is zero for all values of $Q$ smaller than that associated with the first Bragg peak, leading to the result $S(Q \to 0) = 0$. Also the relative variance of the number fluctuations is clearly zero on length scales that are much greater than the size of the unit super-cell. This result is only strictly true in the absence of diffuse scattering at a temperature of absolute zero. The static structure factor of crystals will be finite at finite temperat only holds strictly at absolute zero of temperature, as defects and anharmonic effects mean that the compressibility is non-zero. Note it is important to take the limit $Q \to 0$ so as to avoid the peak at the origin. For all periodic models at large enough length scales (corresponding to small enough $Q$), the static structure factor will go to zero as the static limit is approached, due to the hyperuniformity associated with the crystallinity. Nevertheless we can get meaningful results if we restrict ourselves to distances less than the size of the super-cell, and $Q$ values that are small ($\sim 1/L$ where $L$ is the linear dimension of the supercell) but not too small.

For non-crystalline systems, like amorphous silicon and vitreous silica, we will show that determining the relative variance of $N(R)$ for various sampling radii $R$ and extrapolating the result as $R \to \infty$ provides a much more precise method of extracting the limit $S(Q \to 0)$ from a finite model. Indeed it is the optimal procedure. The relative variance has been thoroughly described by Torquato and Stillinger (Torquato & Stillinger, 2003) and equation (58) from their paper can be written for spherical sampling volumes as

$$\frac{\langle N(R)^2\rangle - \langle N(R)\rangle^2}{\langle N(R)\rangle} = 1 - \rho_0 \frac{4\pi}{3} R^3 + \frac{1}{n}\sum_{i\neq j}^{n} \alpha(r_{ij}; R) \qquad (8)$$

where $n$ is the number of atoms in the model, and the function $\alpha(r_{ij}; R)$ is the fractional intersection volume of two (continuum) spheres, with radii $R$ and centers separated by $r_{ij}$. The function $\alpha(r_{ij}; R)$ is proportional to the probability of two points, separated by $r_{ij}$, both being contained within a randomly placed sphere of radius $R$, and has a form given by Torquato and Stillinger in equation (A11) as

$$\alpha(r; R) = 1 - \frac{3}{4}\frac{r}{R} + \frac{1}{16}\left(\frac{r}{R}\right)^3 = \left(1 - \frac{r}{2R}\right)^2 \left(1 + \frac{r}{4R}\right) \qquad \text{if } r \leq 2R \qquad (9)$$

and zero if $r > 2R$. This is just the shape function that is widely used in describing scattering from spherical micro-crystallites (Lei *et al.*, 2009), but is used in quite a different context here, as it is merely an arbitrary but convenient sampling volume. Using the real-space pair density $\rho(r)$ to convert the sum in equation (8) into an integral, we can write

$$\frac{\langle N(R)^2\rangle - \langle N(R)\rangle^2}{\langle N(R)\rangle} = 1 - \rho_0 \frac{4\pi}{3} R^3 + \int_0^\infty 4\pi r^2 \rho(r) \alpha(r; R) dr. \qquad (10)$$

Using the identity

$$\rho_0 \frac{4\pi}{3} R^3 = \rho_0 \int_0^\infty 4\pi r^2 \alpha(r; R) dr \qquad (11)$$

we obtain the following result

$$\frac{\langle N(R)^2\rangle - \langle N(R)\rangle^2}{\langle N(R)\rangle} = 1 + \int_0^\infty 4\pi r^2 [\rho(r) - \rho_0] \alpha(r; R) dr \qquad (12)$$

which can be conveniently re-written as

$$\frac{\langle N(R)^2\rangle - \langle N(R)\rangle^2}{\langle N(R)\rangle} = 1 + \int_0^\infty r G(r) \alpha(r; R) dr. \qquad (13)$$

Comparing equation (13) to equation (7), they are clearly equivalent as $R \to \infty$ and $Q \to 0$, as the integrand in equation (13) contains $\alpha(r; R)$ which tends to unity for all $r$ as $R \to \infty$. The presence of $\alpha(r; R)$ arises due to the finite nature of the sampling volume, and acts as a natural convergence factor for the integral in equation (7). Notice that the relative variance of $N(R)$ is *not* related to $S(Q)$ except in the limit as both $R \to \infty$ and $Q \to 0$. The sampling volume factor $\alpha(r; R)$ for a sphere can be written as a Taylor expansion in integer powers of $1/R$, allowing the relative variance to be written in the form

$$\frac{\langle N(R)^2 \rangle - \langle N(R) \rangle^2}{\langle N(R) \rangle} = a + b/R + O(1/R^2) \qquad (14)$$

where $a = S(0)$ describes the volume dependence, and $b$ describes the surface dependence associated with the sampling volume. In conjunction with equation (2), equation (14) is therefore a simple but exact relation that allows one to obtain the static structure factor $S(Q \to 0)$ from a large model structure, contained within a super-cell that periodically repeats, and avoids problems associated with extrapolating an oscillating function. We have found this to be the best possible procedure.

## 3. Results

### 3.1. Amorphous silicon models

One major focus of this paper is to determine the limit $S(Q \to 0)$ for amorphous silicon from computer models which serves as a prediction for this important material. As discussed earlier, there is more than one way to find the limit $S(Q \to 0)$, and we will explain the numerical results obtained with all of them here.

The first approach is shown in Fig. 1, where we show the most direct calculation of $S(Q)$ using equation (6) at the points $Q_{lmn} = \frac{2\pi}{L}\sqrt{l^2 + m^2 + n^2}$ determined by the super-cell. While this gives a good overall description of $S(Q)$, it is very limited at small $Q$ and extrapolation or analytic continuation to $Q = 0$ is not possible, even for the much larger 100K model. This is because the finite size oscillations are too severe. Note that the higher density of the 100K model leads to a shift of the peaks to slightly larger $Q$ values. Note also that the structure factor approaches unity at large $Q$ as it must, which sets the scale for comparison for the limit $S(Q \to 0)$. No harmonic phonons (or zero point motion) were added to any of the results in this paper. The inclusion of phonons would have the effect of adding a term that goes as $Q^2$ at small Q, but this would vanish as $Q \to 0$.

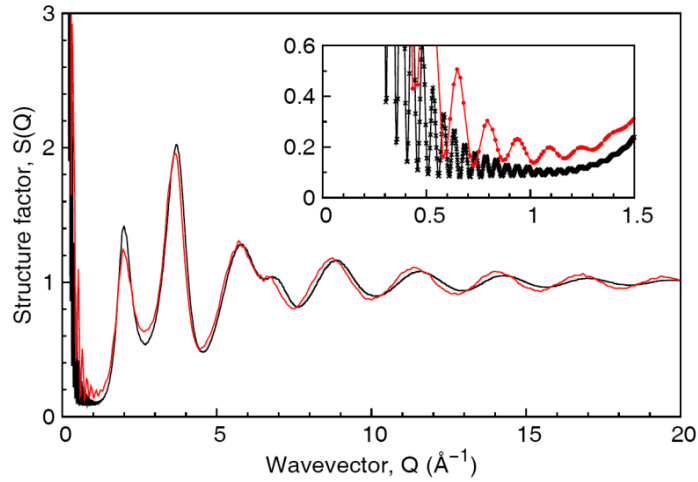

***Figure 1*** *(Color online) The structure factor for amorphous silicon is calculated directly using equation (6) at the super-cell values $Q_{lmn}$, shown in the inset as red circles for the 4096 atom model and black crosses for the 100K model.*

The second method is the Fourier transform method in which $S(Q)$ is determined from the sine transform using equation (1) with $G(r)$ input from the model. Both models of amorphous silicon, with 4096 and 100K atoms are used in Fig. 2 which shows the distribution $G(r) = 4\pi r[\rho(r) - \rho_0]$. Notice the differences in the two silicon models. The difference of 5% in the densities is apparent at small $r$ where $G(r) = -4\pi r\rho_0$, and by the small shift in the peak positions. For comparison, the average separation of bonded silicon atoms determined from the first peak is 2.35Å in the 4096 atom model but only 2.31Å in the 100K model. An isotropic contraction of the whole system does not affect the limit $S(Q \to 0)$, so to first order, the overly dense 100K model should give appropriate values in the limit, as there is no length metric in the limit $Q \to 0$.

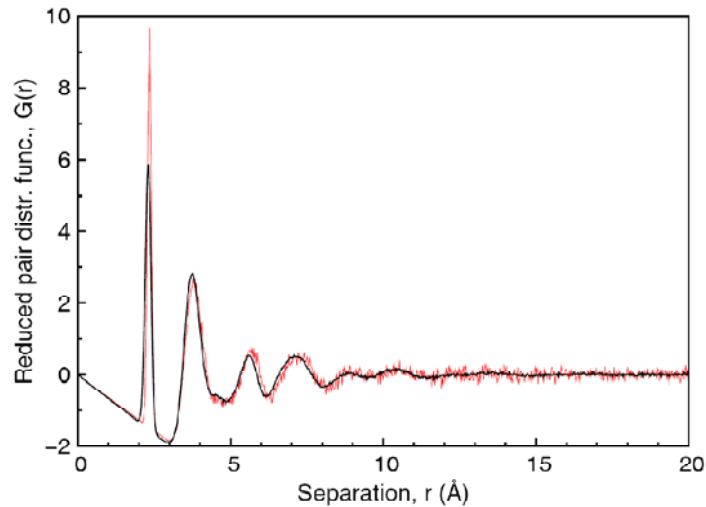

*Figure 2* *(Color online) The pair distribution function $G(r)$ for amorphous silicon for the 4096 atom model (rough red line) and the 100K model (smooth black line).*

The structure factor can be found by applying equation (1) using $G(r)$ for each model. Only the structure factor of the 100K model is shown in Fig. 3, where, even here, the difficulty of trying to extrapolate to $Q = 0$ is again apparent, although the situation is improved somewhat from the direct method shown in Fig. 1.

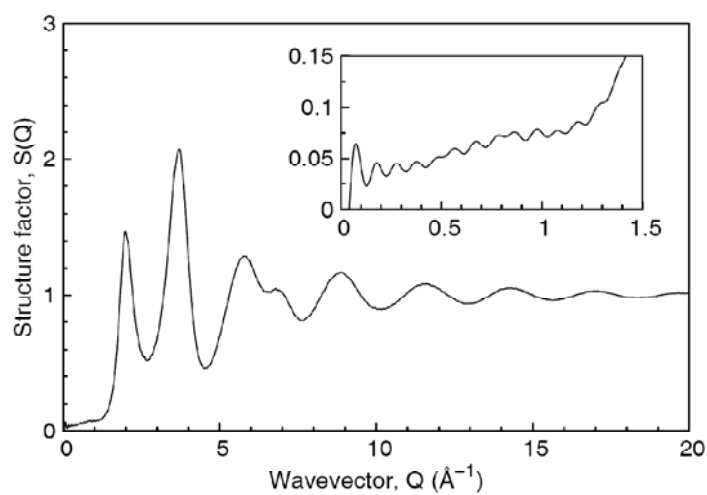

*Figure 3* *The structure factor $S(Q)$ for amorphous silicon obtained directly from equation (1) for the 100K model. The insert shows the small $Q$ region expanded.*

From the inset of Fig. 3 that displays $S(Q)$ at small $Q$, the structure factor of the 100K model still displays significant oscillations due to finite size effects. Of course these oscillations are even more pronounced for the 4096 atom model, which is not shown. These effects arise from the truncation of $G(r)$ beyond $L/2$ (half the width of the cubic super-cell), beyond which $G(r)$ is almost but not quite zero. The source of the oscillations is apparent from their wavelength of $2\pi/(L/2)$. A very approximate limit of $S(Q \rightarrow 0) \cong 0.03$ can be extrapolated by eye for the 100K model from Fig. 3, through the ripples in the insert, but the uncertainty is almost as large as the value itself. For the smaller 4096 atom model, the oscillations are even larger, making any attempt to extrapolate $S(Q)$ quite hopeless. Smoothing techniques can be used to attenuate the oscillations, but are not very convincing. There is a better approach.

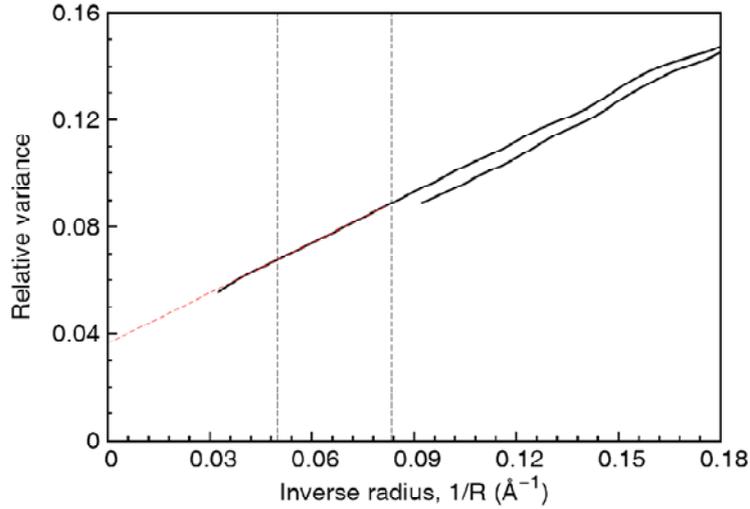

*Figure 4* (Color online) The relative variance in the number fluctuations in amorphous silicon is computed within spheres of various radii R using the sampling volume method. The extrapolated value of S(0), which is just the limit of the relative variance for small 1/R, is given by $S(Q \to 0) = 0.035 \pm 0.001$ *for the 100K mode using equation (2). The vertical dashed lines indicate the range over which the linear fit was performed. It can be seen that the smallest value of 1/R for the 4096 atom model is larger than the upper limit of the range over which the relative variance is linear and therefore a reliable extrapolation cannot be made.*

An alternative to the Fourier transform approach derived in Section 2.3 involves finding the relative variance within finite sampling volumes of increasing size (but identical shape- we have used spheres) and extrapolating to the thermodynamic limit. This has the great operational advantage of avoiding oscillations. The relative variance in the number of atoms within spheres of different radii is plotted in Fig. 4 for both silicon models. The distribution $G(r)$ can only be computed safely out to $r = L/2$ due to the periodic nature of the model. As the sampling volume factor $\alpha(r; R)$ for a sphere is non-zero out to $r = 2R$, the relative variance should only be computed using equation (13) out to $R = L/4$, causing the curve for the 4096 model to terminate at a larger value of $1/R = 4/L$ than that for the 100K model. The relative variance for the 100K model shows a definite linear region within the interval $12\text{Å} < R < 20\text{Å}$ or $0.05\text{Å}^{-1} < 1/R < 0.083\text{Å}^{-1}$. From Fig. 2, the lower limit $R_{min} = 12\text{Å}$ corresponds to the distance at which strong correlations in atom pair separations all but vanish. The upper limit $R_{max} = 20\text{Å}$ corresponds to the radius at which the relative variance within the spherical volumes begins to deviate noticeably from its linear behaviour due to the finite size of the periodic model. The maximum possible radius given the sampling volume argument above is $L/4 = 31\text{Å}$, so $20\text{Å} \approx L/6$ represents a conservative and safe cut-off. If the largest sampling volume for which the relative variance maintains linear behaviour is assumed to be determined by the ratio of the width of the sampling volume to the width of the model, we would expect the linear region to be entirely absent for the 4096 atom model, as $R_{max} = L/6 \cong 7.2$Å is less than the lower limit $R_{min} = 12$Å. Indeed this is what is observed in Fig. 4 for the 4096 atom model, as the oscillations at large values of $1/R$ are still significant by the time the lower limit of $4/L$ is reached. These observations would imply that there is a critical size that a model should be in order for a good extrapolation to $S(Q \to 0)$ in the thermodynamic limit to be possible. At a bare minimum, the width of the box (or for general shapes, the minimum diameter) should be greater than six times the distance over which strong correlations in atom pair separations persist in order for a linear fitting window to exist. For amorphous silicon, this bare minimum would correspond to a periodic super-cell with sides of length 70Å containing ~18,000 atoms. To get a window of decent size for the linear fit, it would be very difficult to work with a model of less than ~50,000 atoms. Triple this amount, ~150,000 atoms, is needed for vitreous silica.

The value of the limit $S(Q \to 0)$ found from linear extrapolation over the linear region of the 100K model is $S(Q \to 0) = 0.035 \pm 0.001$, where the uncertainty represents the spread in the values of the intercept that result for different choices of the fitting interval. An experimental determination of this number would be very interesting, plus the temperature dependence (see comments relating to vitreous silica in the next section).

### 3.2. Vitreous silica model

The behaviour of density fluctuations for the 100K model of amorphous silicon can be compared to a very large model of vitreous silica (300K model) produced by the same group (Vink & Barkema, 2003).

In general for polyatomic systems, it is useful to define partial pair distribution functions (PPDFs) and their corresponding Faber-Ziman partial structure factors (Faber & Ziman, 1965). For vitreous $SiO_2$, the three PPDFs are $G_{SiSi}(r)$, $G_{OO}(r)$, and $G_{SiO}(r)$, where the PPDFs are computed using the subsets of atom types specified by their respective subscripts. Fig. 4 displays the PPDF $G_{SiSi}(r)$ superimposed on $G(r)$ from the 100K silicon model, where the silicon distances in the 300K model have been decreased by a factor of 1.33 to make the silicon atom densities the same.

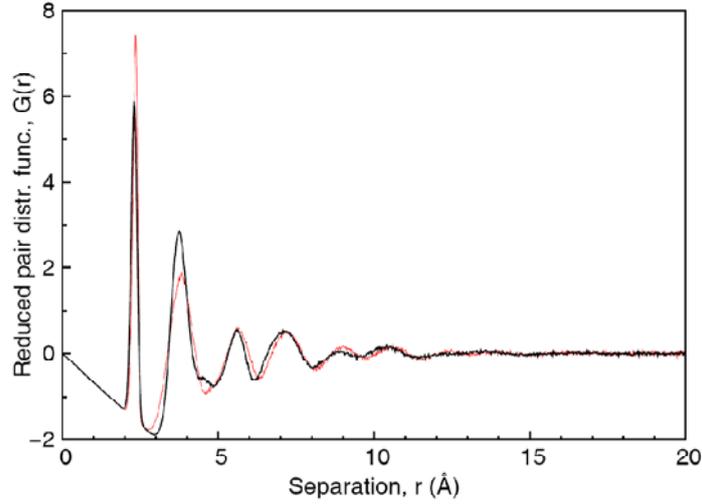

*Figure 5* (Color online) The pair distribution function $G_{SiSi}(r)$ for the 300K vitreous silica model (thin red) rescaled by a length factor of 1/1.33 and superimposed on the same distribution from the 100K amorphous silicon model (thick black, as in Fig. 2).

Using the *rescaled* PPDF $G_{SiSi}(r)$ of vitreous silica as an example of a highly distorted model for amorphous silicon leads to $S_{SiSi}(Q \rightarrow 0) = 0.039 \pm 0.001$ by applying the volume sampling method, and is remarkably close to the value of $S_{SiSi}(Q \rightarrow 0) = 0.035 \pm 0.001$ for the 100K model obtained in the previous section. Thus it appears that the fourfold tetrahedral coordination of the amorphous network is the most important factor is determining $S(Q \rightarrow 0)$ for amorphous silicon.

The three associated partial structure factors $S'_{SiSi}(r)$, $S'_{OO}(r)$, and $S'_{SiO}(r)$ can be found from their respective PPDFs through the sine Fourier transform (Salmon, 2007)

$$Q[S'_{\alpha\beta}(Q) - 1] = \rho_{tot} \int_0^\infty 4\pi r [g_{\alpha\beta}(r) - 1] \sin Qr \, dr \qquad (15)$$

where $\rho_{tot}$ is the number density associated with *all* the atoms in the system, $g(r)$ is the reduced pair distribution function, a scaled version of $\rho(r)$ such that it oscillates about unity at large $r$, and $\alpha$ and $\beta$ define the atom pairs used in the distribution function. This definition of the partial structure factor differs from the intuitive definition that would be obtained if atoms of each type were isolated. This more intuitive definition (for which we use unprimed notation) is represented by partial structure factors of the form

$$Q[S_{\alpha\alpha}(Q) - 1] = \rho_\alpha \int_0^\infty 4\pi r [g_{\alpha\alpha}(r) - 1] \sin Qr \, dr \, . \qquad (16)$$

These two distributions are simply related by

$$S_{\alpha\alpha}(Q) - 1 = (\rho_\alpha/\rho_{tot})[S'_{\alpha\alpha}(Q) - 1] = c_\alpha[S'_{\alpha\alpha}(Q) - 1] \qquad (17)$$

where $c_\alpha = \rho_\alpha/\rho_{tot}$ is the fraction of atoms of type $\alpha$. Three Bhatia-Thornton structure factors (Bhatia & Thornton, 1970, Salmon, 2006, 2007, Fischer *et al.*, 2006) that describe correlations between atom number and concentration can be defined in terms of the three $S'_{\alpha\beta}(Q)$ according to

$$S'_{NN}(Q) = c_{Si}^2 S'_{SiSi}(Q) + c_O^2 S'_{OO}(Q) + 2c_{Si}c_O S'_{SiO}(Q) \qquad (18a)$$
$$S'_{CC}(Q) = c_{Si}c_O[1 + c_{Si}c_O(S'_{SiSi}(Q) + S'_{OO}(Q) - 2S'_{SiO}(Q))] \qquad (18b)$$
$$S'_{NC}(Q) = c_{Si}c_O[c_{Si}(S'_{SiSi}(Q) - S'_{SiO}(Q)) - c_O(S'_{OO}(Q) - S'_{SiO}(Q))] . \qquad (18c)$$

Three of the six unknowns in equations (18) can be found in the limit as $Q \to 0$ by applying the sampling volume method [equation (13)] to $G_{SiSi}(r)$, $G_{OO}(r)$, and $G_{NN}(r)$ (avoiding terms of type $G_{\alpha\beta}(r)$ with $\alpha \neq \beta$). Using the same fitting interval as that for the silicon model results in the limiting values $S_{SiSi}(Q \to 0) = 0.039 \pm 0.001$, $S_{OO}(Q \to 0) = 0.078 \pm 0.002$, and $S_{NN}(Q \to 0) = 0.116 \pm 0.003$, as shown in Fig. 6. Inserting these values into the three Bhatia-Thornton relations (18) and solving for the remaining three unknowns, one finds $S_{SiO}(Q \to 0) = 1.116$, $S_{CC}(Q \to 0) = -1.5 \times 10^{-5}$, and $S_{NC}(Q \to 0) = 0.96 \times 10^{-5}$. Within the uncertainty of the extrapolation, and remembering that there are ~$10^5$ atoms in the model, the limits of the last two Bhatia-Thornton structure factors are consistent with zero, i.e. $S_{CC}(Q \to 0) = S_{NC}(Q \to 0) = 0$. This reflects the fact that the chemical disorder is virtually zero, as only several out of the 100,000 silicon atoms in the model are bonded to another silicon atom instead of to an oxygen atom.

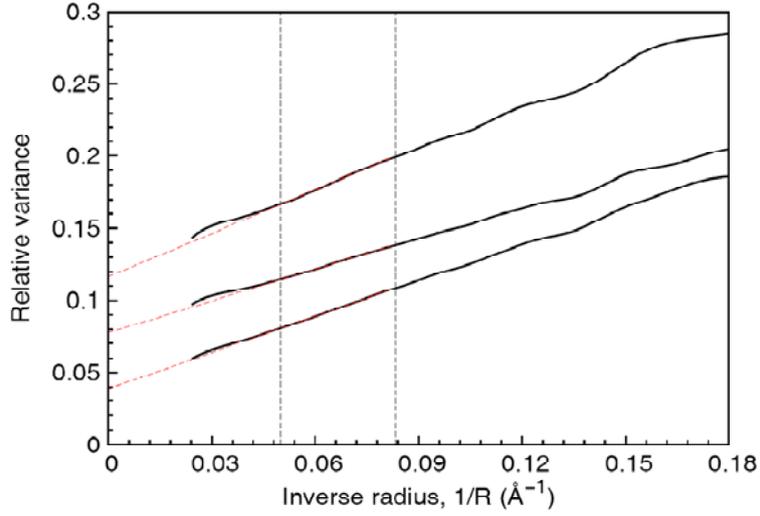

*Figure 6* (Color online) *The relative variance of the number fluctuations in vitreous silica within sampling spheres of radii R. The variance is computed using the sampling volume method and plotted against 1/R. The extrapolated values of S(0), which are just the limits of the relative variances of the number fluctuations for small 1/R, are given by $S_{SiSi}(Q \to 0) = 0.039 \pm 0.001$, $S_{OO}(Q \to 0) = 0.078 \pm 0.002$, and $S_{NN}(Q \to 0) = 0.116 \pm 0.003$. The position and size of the sampling window is determined in a similar way to that described for amorphous silicon.*

If the two quantities $S_{CC}(Q \to 0)$ and $S_{NC}(Q \to 0)$ are *exactly* zero, which we will take to be true from now on, the relationship between the limiting values of the other structure factors simplify greatly, and can all be expressed in terms of a single structure factor rather than the original three. Equations (18a-c) can be rewritten as

$$S'_{SiSi}(Q \to 0) = S'_{NN}(Q \to 0) - \frac{c_O}{c_{Si}} \qquad (19a)$$

$$S'_{OO}(Q \to 0) = S'_{NN}(Q \to 0) - \frac{c_{Si}}{c_O} \qquad (19b)$$

$$S'_{SiO}(Q \to 0) = S'_{NN}(Q \to 0) + 1 \,. \qquad (19c)$$

From equation (17), one can write down the relations

$$S'_{NN}(Q) = S_{NN}(Q) \qquad (20a)$$

$$S'_{SiSi}(Q) = \frac{1}{c_{Si}} S_{SiSi}(Q) - \frac{c_O}{c_{Si}} \qquad (20b)$$

$$S'_{OO}(Q) = \frac{1}{c_O} S_{OO}(Q) - \frac{c_{Si}}{c_O} \,. \qquad (20c)$$

In the thermodynamic limit, the previous six equations relate the limiting values of the seven structure factors, and thus there is only *one* independent quantity. If the independent quantity is chosen to be $S_{NN}(Q \to 0) \equiv S(0)$, the limiting value of the other structure factors that one would find if each atom type *was taken in isolation* can be expressed along with the Bhatia-Thornton number correlation as

$$S_{NN}(Q \to 0) = S(0) \qquad (21a)$$
$$S_{SiSi}(Q \to 0) = c_{Si}S(0) \qquad (21b)$$
$$S_{OO}(Q \to 0) = c_O S(0). \qquad (21c)$$

The scaling factors that exist between these three values when there is no chemical disorder in the system explains why the values found from the sampling volume method follow a 1:2:3 ratio [$S_{SiSi}(Q \to 0) = 0.039 \pm 0.001$, $S_{OO}(Q \to 0) = 0.078 \pm 0.002$, and $S_{NN}(Q \to 0) = 0.116 \pm 0.003$], as $c_{Si} = 1/3$ and $c_{Si} = 2/3$. Notice that this scaling is only present as $Q \to 0$ and of course is not true at a general $Q$. All the analysis of the 300K vitreous silica model can therefore be summarized in a single number by there being virtually no chemical disorder and $S_{NN}(Q \to 0) = S(0) = 0.116 \pm 0.003$.

The expression for the limiting value of the differential scattering cross-section per atom obtained from scattering experiments also simplifies if no chemical disorder is present. The general form of differential cross-section per atom (Fischer *et al.*, 2006, Salmon, 2006, 2007), namely

$$\sum_{\alpha\beta} c_\alpha c_\beta b_\alpha b_\beta \left[ S'_{\alpha\beta}(Q) - 1 \right] + \sum_\alpha c_\alpha b_\alpha^2 \qquad (22)$$

where $b_\alpha$ is the scattering length of atoms of type $\alpha$, can be simplified in the limit $Q \to 0$ by writing the three partial structure factors $S'_{SiSi}(Q \to 0)$, $S'_{OO}(Q \to 0)$, and $S'_{SiO}(Q \to 0)$ in equation (22) in terms of $S_{NN}(Q \to 0) = S(0)$ using equations (20a-c) and (21a-c). Performing the substitutions, one finds that differential cross-section per atom simplifies to

$$[c_{Si}b_{Si} + c_{Si}b_{Si}]^2 S(0) . \qquad (23)$$

Equation (23) is often used to interpret experimental data (Levelut *et al.*, 2002, Levelut *et al.*, 2005, Levelut *et al.*, 2007, Wright *et al.*, 2005, Wright, 2008) under the assumption that the AX$_2$ units can be considered as the basic entity, with an associated scattering factor ($c_{Si}b_{Si} + c_{Si}b_{Si}$). It was not clear to us until doing the present analysis that this was justified, as two out of the four neighboring X atoms are arbitrarily associated with an A atom, and in addition, this AX$_2$ unit may straddle the perimeter of the sampling volume, leading to partial counting.

Nevertheless, the above derivation shows that this widely used phenomenological assumption (23) is indeed justified and correct, subject to there being no chemical concentration fluctuations, so that each A atom is bonded to four X atoms and each X atom is bonded to two A atoms.

Experiments to determine the absolute value of $S(0)$ are not easy because the scattering has to be normalized to a standard, and also because of multiple scattering corrections that are best determined by measuring a number of samples of varying thickness and extrapolating to zero thickness. This complicated procedure has been done recently by Wright (Wright et al., 2005, Wright, 2008), who using equation (23) obtains a value for vitreous silica of $0.0300 \pm 0.0016$ per formula unit, which by incorporating the factor of three leads to a value for the static structure factor of $S(0) = 0.0900 \pm 0.0048$. Note that Wright was able to get down to $Q \approx 0.02 \text{Å}^{-1}$, which is about a factor of 10 better than can be obtained with the 300K model. The model value of $S(0) = 0.116$ is about 20% higher than the experimental value, which we comment on below. Nevertheless, this is the first calculation of $S(0)$ from a model of vitreous silica and is gratifyingly close to the experimental value.

We note that Salmon (Salmon, 2006, 2007, Salmon et al., 2007) has made measurements of structure factors on a number of $AX_2$ glasses using isotopes so that the partial structure factors can be found, and hence $S_{NN}(Q)$. These experiments are a real tour de force but not specifically designed to measure the $Q \rightarrow 0$ limit. Not being performed at very small $Q$ (down to $Q \approx 0.5 \text{Å}^{-1}$) and they are only indicative, but approximate values can be extrapolated from the plots of the partial structure factors at small $Q$, giving values between ~ 0.1 and ~ 0.15 for Ge0$_2$, GeSe$_2$, and ZnCl$_2$ (Salmon, 2006, 2007, Salmon et al., 2007). These are very close to the more accurate value for vitreous silica obtained by Wright et al. and to the model calculation, suggesting *perhaps* that this value, $S(0)$ ~ 0.10 is a general feature of $AX_2$ glasses, as a value ~ 0.035 is characteristic of single component tetrahedral glasses.

## 4. Comments

For a system in thermal equilibrium, like a liquid, we expect equation (4) to hold. It is useful to use this relation to access how far amorphous silicon, as well other amorphous materials and glasses, are from equilibrium. The compressibility $\chi_T$ of amorphous silicon is between 2 x 10$^{-11}$ m$^2$/N and 3 x 10$^{-11}$ m$^2$/N, obtained from silicon-aluminum alloy data extrapolated to zero aluminum doping (Keita & Steinemann, 1978). Using $\rho_0 = 0.05$ atoms/ Å$^3$ (Custer et al., 1994, Laaziri et al., 1999), and using room temperature of 300K, we find from equation (4) that $0.004 < S(0) < 0.006$, which is an order of magnitude less than the computer model value of 0.035. If we use the *melting temperature* of crystalline silicon of roughly $T = 1685K$ (Grimaldi *et al.*, 1991), this estimate increases to $0.023 < S(0) < 0.035$, where we note that both the density $\rho_0$ and the compressibility $\chi_T$ are only weakly dependent on temperature so that almost all of the temperature dependence in equation (4) comes through the temperature factor $T$ itself. Nevertheless, the figures based on high temperatures are in the general area of the value of $S(0) = 0.035$ determined from the 100K model, which is not unreasonable. Note that the comparison is a little less favourable if we use the melting temperatures of 1220K to 1420K for amorphous silicon (Donovan *et al.*, 1989, Grimaldi *et al.*, 1991), which leads to $0.017 < S(0) < 0.030$.

The most extensive data on the static structure factor for liquid and vitreous silica has been assembled by Levelut and co-workers (Levelut *et al.*, 2002, Levelut *et al.*, 2005, Levelut *et al.*, 2007). They have used small angle x-ray scattering, with wavevectors $Q$ down to $\approx 0.027 \text{Å}^{-1}$, which is comparable to that obtained from the 300K vitreous silica model used in this paper. However their values do seem systematically low in not only the glass phase, but also in the liquid phase where we would expect equation (4) to hold. Note that there is a factor of $900 = (30)^2$ between the data of Wright and Levelut, due to the electron units used by Levelut, which in turn differs by a factor of three from the conventional definition of the structure factor as used here and by Salmon (Fischer *et al.*, 2006, Salmon, 2006, 2007).

To try and gain some perspective, we have used existing compressibility measurements (Bucaro & Dardy, 1976, Wright *et al.*, 2005, Wright, 2008) and assumed equation (4) to be true in order to renormalize the Levelut data upward by a factor of 1.43, which is now re-plotted in Fig. 7. This scale factor is the ratio of the liquid compressibility value quoted by Bucaro (Bucaro & Dardy, 1976) to the average of the two liquid compressibility values quoted by Levelut (Levelut *et al.*, 2005), i.e. $1.43 = 8.50/[(6.16+5.69)/2]$.

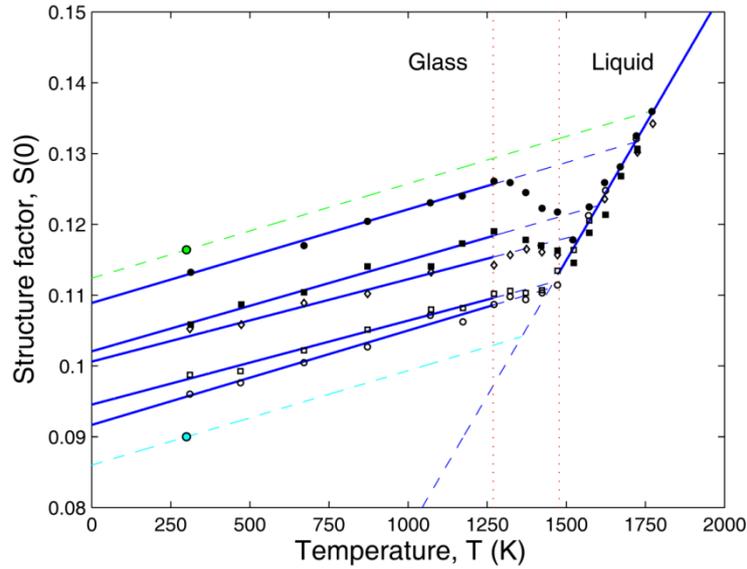

***Figure 7*** *(Colour online) The points and fitted blue solid lines in both the glass and liquid region of silica are digitized from Fig. 2 of Levelut (Levelut et al., 2005) multiplied by a factor of 1.43 as described in the text. The five lines in the glass phase correspond to fictive temperatures of 1373K (open circles), 1473K (open squares), 1533K (solid squares), 1573K (open diamonds), and 1773K (solid squares). The lower isolated point (cian) is from Wright (Wright et al., 2005) and the upper isolated point (green) is from the computer model used in this paper.*

Fig. 7 raises many interesting questions relating to glass structure and the fictive temperature (Geissberger & Galeener, 1983). It is clear from the data of Levelut et al. that the fictive temperature is very close to where the extrapolated straight lines from the glass phase intersect with the liquid structure factor. Note that the temperature dependence is considerably lower in the glass phase and is due to the thermal vibrations about a fixed network topology (Weinberg, 1963, Wright *et al.*, 2005, Wright, 2008). A most important and intriguing question is *how is information about the fictive temperature embedded in the glass at room temperature?* The information presumably involves ring statistics and possibly the oxygen angle distribution, but it is subtle and will require careful modelling to resolve. All models used will have to be as large as those used in this paper to get reliable values for $S(0)$, as discussed earlier. The dashed lines drawn through the two isolated points in Fig. 7, parallel to the Levelut et al. lines, suggest a fictive temperature of ~1360K for the Wright sample and a fictive temperature of ~1780K for the 300K vitreous silica model of Vink and Barkema (Vink & Barkema, 2003), which is close to the value of 1740K used for the start of the quench in their computer model. Note that while this close agreement is promising, one must not forget that the computer model is quenched at a much more rapid rate than an actual sample, and it is not clear how close the values of the experimental temperature and the "computer" temperature should be. One might argue that the quench rate is of secondary importance to the fictive temperature in determining the glass structure, but this is very speculative and requires further study.

## 5. Concluding remarks

The static structure factor $S(Q \to 0)$ for two non-crystalline materials, amorphous silicon and vitreous silica, lies between that of a crystalline solid (where it is close to zero) and that of liquid vitreous silica. From the computer model of Mosseau, Barkema and Vink, the static structure factor for amorphous silicon is computed to be $S(0) = 0.035 \pm 0.001$. This non-zero value is caused by density fluctuations, similar to those found in a liquid, even though the system is far from thermal equilibrium, and seems to be determined largely by the tetrahedral coordination in the amorphous material. This result awaits experimental confirmation, for which it will also be interesting to measure the temperature dependence, caused by thermal fluctuations about the network structure.

For vitreous silica, the situation is richer as the results depend on both the actual temperature and the fictive temperature, as demonstrated clearly by the experimental results of Levelut et al. The large periodic computer model of Vink and Barkema gives a reasonable value $S(0) = 0.116 \pm 0.003$ for vitreous silica at room temperature which corresponds to an experimental fictive temperature of about 1780K, close to 1740K used computationally to achieve the quenched structure. The intriguing question that remains unanswered is how the information about the fictive temperature is encoded within the network structure, and we speculate that it is in the distinct ring statistics, but this remains to be seen.

We should like to acknowledge very useful discussions with Austen Angell, Neville Greaves, Gabrielle Long, Simon Moss, Phil Salmon, Adrian Wright and last but not least Paul Steinhardt, who inspired this work by asking what the static structure factor was in amorphous silicon. We also to Normand Mouseau, Gerard Barkema and Richard Vink, for the coordinates of their large amorphous silicon and vitreous silica models, which made this work possible. This work was supported as by of FRG effort funded by NSF DMR 0703973.